\documentclass{elsart}

\usepackage{graphicx}

\usepackage{amssymb}

\begin{document}

\begin{frontmatter}



\title{
Beam Test of a Prototype Detector Array for the PoGO
Astronomical Hard X-Ray/Soft Gamma-Ray Polarimeter
}


\author{T. Mizuno\corauthref{cor}},
\corauth[cor]{
Corresponding author. 
Tel.:+1-650-926-2595,
fax.:+1-650-926-4979. \\
{\it Email address:} mizuno@SLAC.Stanford.EDU}
\author{T. Kamae, J. S. T. Ng and H. Tajima}
\address{Stanford Linear Accelerator Center,
Menlo Park, California, 94025, USA}

\author{J. W. Mitchell and R. Streitmatter}
\address{NASA Goddard Space Flight Center, 
Greenbelt, Maryland, 20771, USA}

\author{R. C. Fernholz and E. Groth}
\address{Princeton University, Princeton, New Jersey, 08544, USA}

\author{Y. Fukazawa}
\address{Hiroshima University, Higashi-Hiroshima, 
Hiroshima, 739-8526, Japan}

\address{}

\begin{abstract}
Polarization measurements
in the X-ray and gamma-ray energy range 
can provide crucial information on massive
compact objects such as black holes and neutron stars. 
The Polarized Gamma-ray Observer (PoGO) 
is a new balloon-borne instrument designed to measure 
polarization from astrophysical objects in the 30-100~keV range, 
under development by an international
collaboration with members from United States, Japan, Sweden and France.
To examine PoGO's capability, a beam test 
of a simplified prototype detector array was conducted 
at the Argonne National Laboratory Advanced Photon Source. 
The detector array consisted of seven plastic scintillators,
and was irradiated by polarized photon beams at 60, 73, and 83 keV. 
The data showed a clear polarization signal, with a measured modulation
factor of $0.42 \pm 0.01$.  
This was successfully reproduced 
at the 10\% level by the 
computer simulation package Geant4 
after modifications to its implementation of
polarized Compton/Rayleigh scattering.  Details of the beam test and
the validation of the Geant4 simulations are presented.
\end{abstract}
\end{frontmatter}

\section{Introduction}
\label{Introduction}

Measurements of X-ray and gamma-ray polarization are
expected to yield important information 
on a wide variety of astrophysical sources such as
isolated pulsars, 
jet-dominated active galaxies,
and accreting black holes and neutron stars.
In the astrophysical environment, polarization arises under a variety of
conditions.  Polarization in synchrotron radiation is due to
the electrons gyrating in ordered magnetic fields \cite{Rybicki},
providing information on the properties of magnetic fields around the source.
The absorption cross-section of photons, as they propagate through 
a strong magnetic field, is polarization and energy dependent, making it
possible to investigate the strong-field environment near the surface of 
a neutron star \cite{Alice1,Alice2}.  
Polarization can also result from the Compton scattering of an
incident photon flux in the accretion disks around compact stars
and active galactic nucleus.
In all cases, the orientation of the polarization plane
depends on the orientations of the magnetic fields and the accretion disk,
hence it is a powerful probe of the source geometry.
However, despite its importance, X-ray and gamma-ray polarization
has been measured by only two experiments,
one on the OSO-8 satellite which viewed the Crab at 2.6 and 5.2 keV and 
measured the polarization using Bragg diffraction 
\cite{Weisskopf1976,Weisskopf1978,Silver1978},
the other on the RHESSI satellite which 
has reported the detection of polarization
for a gamma-ray burst \cite{RHESSI2003}.
Thus far, there has been no systematic study of polarization
in high-energy astrophysics due to the lack of sensitivity, and
polarization at hard X-ray and soft gamma-ray energies,
where non-thermal processes are 
likely to produce a
high degree of polarization, is yet to be explored.
We note that, on the other hand,
there has been a long history of attempts to
measure polarization of hard X-rays
in solar flares followed by recent missions of
SPR-N on CORONAS-F \cite{Bogomolov2003} 
and RHESSI \cite{McConnell2004}.
See an introduction section in \cite{McConnell2004} and
references therein for a review.

To carry out these measurements, we are constructing a new
balloon-borne instrument, the Polarized Gamma-ray Observer (PoGO),
which employs coincident detection
of Compton scattering and photo-absorption to measure
polarization in the 30--100 keV energy range.
The PoGO instrument utilizes an adaption 
to polarization measurements of the well-type phoswich counter design 
developed through a series of balloon experiments 
\cite{Gunji1992,Gunji1994,Miyazaki1996,Yamasaki1997,Kamae1992,Kamae1993,Takahashi1993}
and implemented in the ASTRO-E/ASTRO-E2 satellite mission
as the Hard X-ray Detector 
\cite{Kamae1996,Takahashi1996,Tanihata1999,Makishima2001}.
Through these balloon experiments and tests on the ground,
the well-type phoswich counter design has 
been shown to be highly effective in reducing background
and has achieved high sensitivity in measuring hard X-ray spectra
\cite{Gunji1992,Takahashi1993,Kamae1996,Tanihata1999}.
The conceptual design of the PoGO instrument is shown in Figure~1.
A hexagonal array of fast plastic scintillators 
function as a Compton polarimeter
for hard X-rays/soft gamma-rays
by measuring the azimuthal scattering angle asymmetry.
This is surrounded by bottom and side anti-coincidence detectors
made of bismuth germanate oxide (BGO) scintillators.
The aperture is defined by active collimators 
made from tubes of slow plastic scintillator.
A similar modular polarimeter can be found as GRAPE
\cite{GRAPE1999,GRAPE2004}, which is optimized for measuring
polarization of energy $\ge$ 100~keV from
solar flares and gamma-ray bursts.

To simplify readout, the instrument is organized as an array of
hexagonal phoswich units, each consisting of a fast scintillator 
(decay time $\tau \sim 2~{\rm ns}$), 
a slow plastic scintillator active collimator 
($\tau \sim 300~{\rm ns}$), 
and a bottom BGO anti-coincidence detector 
(${\rm \tau \sim 300~ns}$), all viewed by a single photomultiplier tube (PMT).
In order to reduce the background due to downward
atmospheric gammas and cosmic diffuse gammas
coming from outside the field-of-view,
we also use a thin high-Z metal foil, 
wrapped around the active collimator tubes, as 
passive collimators. Signals from the fast plastic scintillator and those
from the slow plastic or BGO scintillator
can be separated using a pulse shape discrimination technique
by examining signal development in two different time windows.

The design parameters of PoGO, 
and its expected performance and response to
polarized gamma-rays 
have been studied in extensive Monte-Carlo simulations
using the EGS4 \cite{EGS4} and Geant4 \cite{Geant4} computer program packages.
The current design of PoGO consists of 217 phoswich units
composed of 3~cm long BGO scintillator,
20~cm long fast scintillator and 60~cm long
slow scintillator, each one is of $\sim$2.6~cm wide.
The width is determined by the availability of PMTs with
large photocathode area, and the length of 
fast scintillators is determined to
maximize the sensitivity for polarization.
Our initial study with Monte-Carlo simulation showed that
the length of 15--20~cm could be optimum:
Scintillators longer than photon attenuation length 
for the energy of our interest are
preferred, since the forward scattered photons are still
highly polarized, 
and hence events with double Compton scatterings
in a scatterer scintillator exhibit
high azimuthal scattering angle asymmetry
(see also a discussion in \S~3).
A geometrical area is ${\rm \sim 930~cm^{2}}$
and an effective area is ${\rm \sim 230~cm^{2}}$ at 40~keV,
when we require that two fast scintillators detect a hit with a
threshold level of 3~keV to select Compton events.
The Monte-Carlo simulations also show that 
PoGO will achieve the sensitivity to measure
better than 10\% polarization from a 100 mCrab source
in a single 6 hour balloon observation.
More details of the instrumentation can be found in
\cite{Chen2003}.

\section{Beam Test at the APS}
\label{Beam Test at the APS}

To validate our simulations of polarized Compton scattering process,
and to demonstrate the ability of PoGO to measure polarization,
we have conducted a test-beam experiment with a PoGO prototype
at the Advanced Photon Source Facility (APS)
of the Argonne National Laboratory during November 10--18, 2003.  
Because the main objective of the beam test was to 
measure the sensitivity, in the energy regime of interest, 
of the PoGO technique of Compton polarimetry, the prototype
only included the fast plastic scintillators.
This simplified the read-out scheme, and we
did not need to use the pulse shape discrimination technique.
The prototype was arranged as an
array of 7 hexagonal Saint-Gobain BC-404 scintillators, 
each 2.68~cm wide and 20~cm long,
as shown in Figure~2 (which includes the numbering scheme).  
Each scintillator was glued directly 
to a Hamamatsu R580 PMT (3.4 cm photocathode diameter,
larger than that of PMTs used for PoGO flight).  
The center scintillator acted as a Compton scattering target 
and the outer six scintillators, 
separated from the center scintillator by 2.2~cm, 
detected the scattered photons.  
In the final PoGO instrument the hexagonal detector units 
will be tightly bundled together 
in a close-packed hexagonal array \cite{Chen2003}.  
However, photons scattered in one detector are likely to pass 
through at least its nearest neighbors before being photo-absorbed 
in other units. Thus, the prototype array approximates 
a region of the final PoGO instrument. 
Tests with slow scintillator collimators and 
bottom/side BGO scintillators and with larger number of units
will be done in fiscal year of 2005 and 2006.

The experiment was installed in the MUCAT 6ID-D station,
which enabled us to control (reduce) the beam intensity
down to an appropriate flux for our instrument.
The test detector was exposed to plane-polarized photon beams
at 60~keV, 73~keV, and 83~keV delivered through an undulator and a
double-scattering monochromator upstream. 
The beam line was operated in a reduced
flux mode by detuning one stage of the monochromater 
with a typical flux of approximately ${\rm 10^7~photons~s^{-1}}$.  
This was further
reduced with attenuators, resulting
in a trigger rate of a few kHz at the experiment.
The degree of polarization was calculated to be 98--99\%; 
there was also the background
(a non-polarized broad-band continuum and photons
not associated with the beam)
which contributed to approximately 2-3\%
of the flux.  

The array was mounted on a rotation stage, 
as shown in Figure~3, to allow measurement of the
modulation factor by rotating about the center scintillator aligned to
the incident photon beam.  
To measure the detector response to various orientations
of the polarization vector, the instrument was
rotated in 15~degree steps covering the
entire azimuthal angle range (0--360~degree).
For each run, we integrated about 100k events.
Data at the three different beam energies allowed for
calibration of the
detector, providing the energy vs. pulse height scale factors
as well as the energy resolution for each sensor.
The calibration data were obtained by irradiating beam
at each detector with a self trigger.
The energy resolution (full width at half maximum)
at 73~keV spanned the range of 25~\% to 33~\% and scaled with the inverse
of the square-root of energy.

The signals from the PMT anodes 
were fed into charge-sensitive preamplifiers
and then filtered and amplified by shaping amplifiers with
$1~{\rm \mu s}$ shaping time.  The outputs of these amplifiers were used for
both trigger generation and spectroscopy.   A Compton scattering 
trigger was generated by a coincidence of hits
in the central scintillator
and in any one of the surrounding scintillators, 
with a threshold of about 0.5~keV and 5~keV, respectively.
The data acquisition system is shown in Figure~4.
Note that the electronic
readout scheme is different
from that of the final PoGO instrument \cite{Chen2003}.

During the experiment, we observed that the trigger efficiency
varied among the peripheral scintillators.
In order to evaluate this,
we collected an unbiased data sample, at 73~keV for the angular range of
0--150~degree in 30~degree steps,
triggered only by the central scintillator (channel~4).
We applied the event selection criteria described in \S~3
and calculated the trigger efficiency as the ratio of
counts of coincidence trigger run to that of 
channel~4 trigger run for each peripheral scintillator.
The relative efficiency was the highest for channel~7
(normalized to 1)
and the lowest for channel~6 ($\sim 0.65$); the efficiencies 
were determined with a statistical error of $\sim$2.5\%.
These were taken into account in the following analysis.

\section{Data Analysis and Simulation}
\label{Data Analysis and Simulation}

A raw data sample at 73~keV taken with coincidence trigger 
at 0-degree rotation is shown in Figure~5 as a scatter-plot of the
energy detected in the central scintillator and the total energy
detected in all seven scintillators.
The strip
structure at ${\rm \sim 10~keV}$ energy deposition in the
central scintillator was due to the trigger inefficiency
mentioned in \S~2.
We can see a clear separation
between events in which Compton scattered photons were photo-absorbed
in the peripheral scintillators
(total energy deposit at about 73~keV) and
those in which the scattered photons escaped
(total energy deposit below 40~keV).
In order to select valid Compton events,
i.e., events in which incident photon was Compton-scattered
in the central scintillator and photo-absorbed in 
only one of the peripheral scintillators,
we applied the following criteria:
1) the central scintillator and only one of the outside scintillators 
detected a hit, where detection threshold was set at 3~keV,
well above the single photo-electron noise
(corresponding to 1--1.5~keV energy deposition);
2) the energy detected in the central scintillator was below 40~keV and less
than half of the total energy deposit, to ensure that 
the central scintillator scattered the incident photon; and
3) the total energy deposit is ${\rm 73.2 \pm 25~keV}$, consistent with
incident beam energy within detector energy resolution.
The selection criteria are also shown in Figure~5.
(Note that the surrounding slow plastic and BGO scintillators
onboard the PoGO flight-instrument will allow us
to eliminate events where the incident photon
did not deposit its full energy in the fast scintillators,
even without any information on the initial energy.)
Polarization could then be determined by the variation in hit rates
in each peripheral scintillator due to the anisotropy 
in the probability of azimuthal
scattering angles of Compton events.

The results, taking variation in trigger efficiencies into account,
are summarized in Figure~6.
There, the average of counts of channels~1 and 7,
that of channels~2 and 6, and that of channels~3 and 5 are plotted
as a function of instrument rotation angle.
We normalized the data to the number of selected events,
since the total photon flux was not recorded.
A clear modulation can be seen,
with the least counts in scintillators 
along the polarization vector
(e.g., channels~3 and 5 for 0-degree rotation) 
and the highest counts in those 
perpendicular to the vector as expected.
We fit the result to a sinusoidal curve and obtained a
modulation factor (${M_{p}}$) from the maximum (${R_{\rm max}}$) 
and minimum (${R_{\rm min}}$) rates 
measured as a function of azimuthal angle by: 
\begin{equation}
M_{p} = \frac{R_{\rm max}-R_{\rm min}}{R_{\rm max}+R_{\rm min}}
\end{equation}
The resulting modulation factor of $0.421 \pm 0.010$ 
from the normalized data, is consistent with the value of $0.423 \pm 0.012$ 
obtained from the unbiased data sample taken with triggers in the
central scintillator only.
Similar event selections applied to the 
60~keV and 83~keV data samples yielded
a modulation factor of
$0.402 \pm 0.011$ and $0.416 \pm 0.010$, respectively,
indicating that the modulation factor is 
almost independent of the beam energy.

Our results were compared with computer simulations using the Geant4
toolkit (version 5.1) \cite{Geant4} with low energy extensions which are
especially important for simulating polarized photon scattering.
Initial simulations gave a modulation factor of
$\sim 0.37$, resulting in an unphysical polarization value
(over 100\%).  This was due to an incorrect implementation of the
polarized Compton and Rayleigh scattering processes in the code.  
As described in
detail in Appendix~A, we improved it and validated it against
another simulation program EGS4 \cite{EGS4,Namito1993,Namito1994}.
Note that Geant4 provides greater flexibility
in simulating complex geometries compared to other codes
such as EGS4;
therefore it is more suitable for studying the response
of a complicated instrument such as PoGO.

The new Geant4 simulator yielded a modulation factor of $0.488 \pm 0.006$
for a fully-polarized 73~keV photon beam and a perfectly aligned PoGO
prototype detector.  By taking into account the degree of polarization 
of the beam and the effect of the background,
we obtained a modulation factor of about 0.47.
The small (10\%) difference from the data
could be accounted for by the uncertainty of the simulation
(see Appendix~A) and possibly by misalignment of the instrument.
Simulations for photon beams at 60~keV and 83~keV yielded consistent
modulation factors of $0.487 \pm 0.006$ and $0.489 \pm 0.008$,
respectively.
The small ($\sim 4$~\%) energy dependence seen in data
could be due to the energy dependence of the trigger efficiency.

Since the PoGO scintillators 
are much longer than the photon attenuation
length at the test beam energies (${\rm \sim 6~cm}$), the
probability of double scattering in one scintillator
is not negligible.
About $\frac{1}{3}$ of the events that passed the
selection criteria showed energy depositions of
more than 15~keV in the central scintillator
(corresponding to a scattering angle 
of $150^{\circ}$), consistent with double scattering.
Even for these events, 
data collected using triggers with central scintillator hit only
yielded a moderately high modulation factor of $0.285 \pm 0.022$,
indicating that our crude event selection
resulted in moderate modulation factors as well as high photon statistics.
This would be an advantage in astrophysical observations with limited
photon statistics.

\section{Summary}
\label{Summary}

We have conducted a beam test experiment on a prototype of
the Polarized Gamma-ray Observer (PoGO), under development for polarization
studies of high-energy astrophysics processes 
in the hard X-ray/soft gamma-ray regime.
The prototype consisted of seven plastic scintillators read out with PMTs,
and was exposed to polarized photon beams at 60~keV, 73~keV and 83~keV.
We obtained a clear modulation signal 
with a measured modulation factor of $0.42 \pm 0.01$.
We also compared our results with the Geant4 simulation program and
found that its implementation of 
polarized Compton/Rayleigh scattering processes required modifications.
The modified Geant4 simulation results agreed with our data to
10\%.
The new Geant4 was also validated against EGS4
at the 2--3\% level.

\section{Acknowledgement}

This work was partially supported by the
U.S. Department of Energy under contract
DE-AC03-76SF00515 and NASA under
RTOP 187-01-01-07.
We would like to thank the APS staff for their generous and friendly
support, in particular D. Robinson whose help made possible the success 
of this experiment.  
We are also grateful to members of PoGO collaboration.
In particular, we would like to thank
J. Kataoka, Y. Yatsu, T. Ikagawa, 
Y. Yamashita and M .Suhonen for their support for the experiment
at Spring-8, which was indispensable for the success of
this experiment.
We would also like to thank D. Marlow, G. Bogaert,
T. Thurston, A. Scholz and R. Rogers for their support
for the test preparation.
Use of Advanced Photon Source was supported by the U. S.
Department of Energy, Basic Energy Science, Office of Science,
under Contract No. W-31-109-Eng38.

\clearpage

\begin{figure}
\noindent\includegraphics[width=\textwidth]{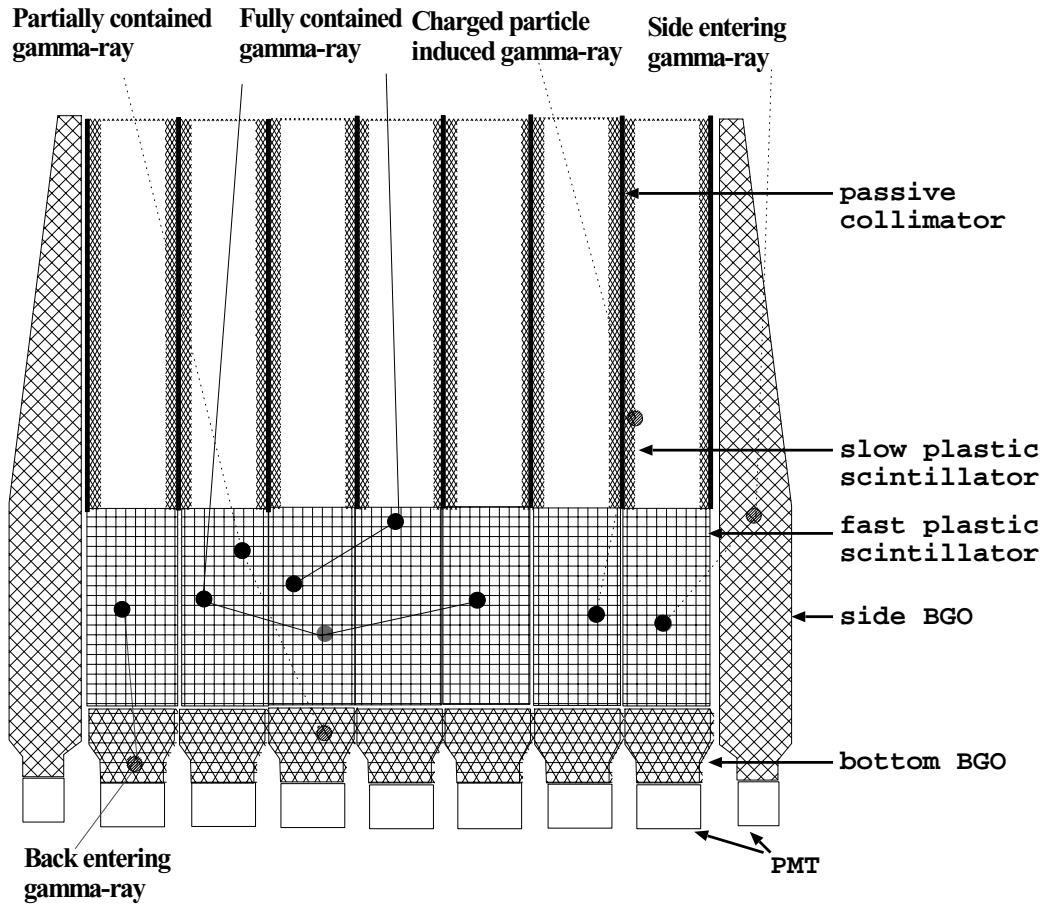}
\caption{
Conceptual design of PoGO.
It consists of an array of well-type phoswich detector units,
each made of a fast plastic scintillator,
a slow plastic scintillator tube,
a thin high-Z metal foil
and a bottom BGO.
A set of side anti-coincidence detectors made of BGO surrounds the
array of phoswich units.
In the figure, representative passages of gamma-rays
are shown with energy deposition marked by circles.
The trigger scheme accepts only the ones marked as
``Fully contained gamma-ray''.
}
\end{figure}

\begin{figure}
\noindent\includegraphics[width=\textwidth]{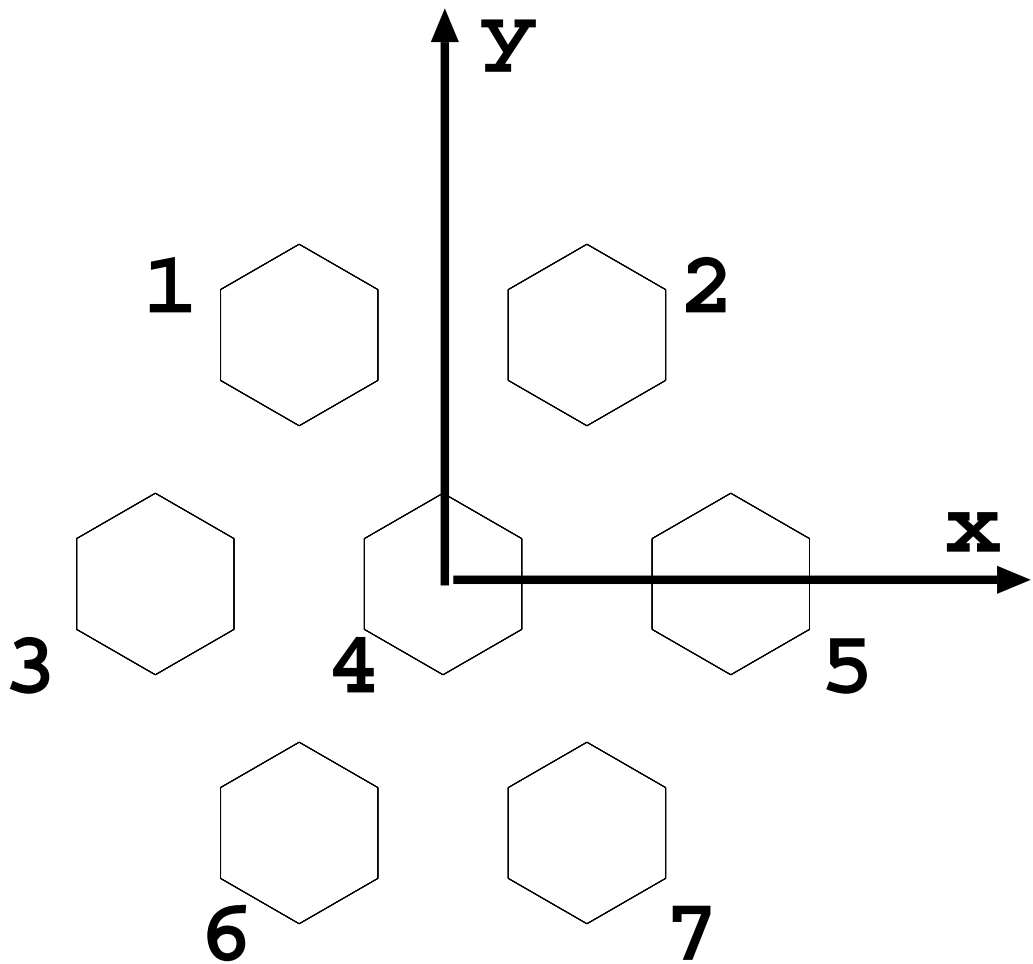}
\caption{
The layout and numbering scheme of scintillators viewed from the
beam origin.
Detector rotation angle is defined to be $0^{\circ}$
when scintillators channels 3, 4 and 5 are along the 
horizontal (x-axis),
and to be $30^{\circ}$ when channels 1, 4 anc 7 are
along the vertical (y-axis).
}
\end{figure}

\begin{figure}
\noindent\includegraphics[width=\textwidth]{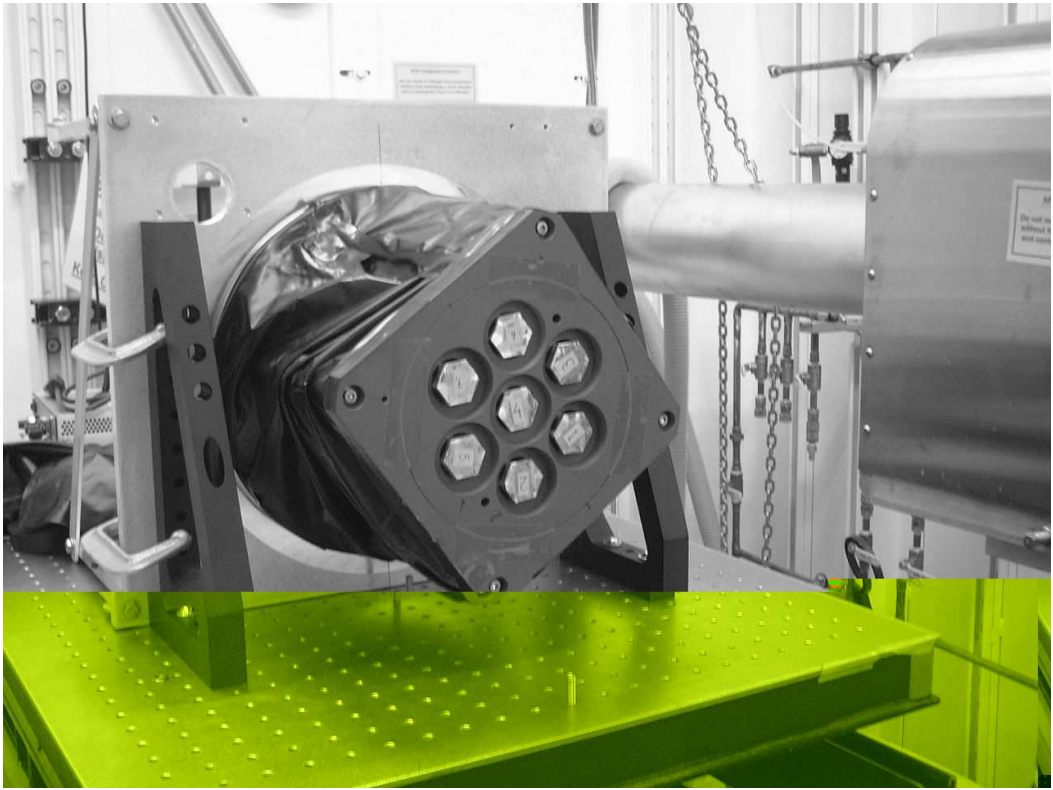}
\caption{
A photograph of the PoGO prototype 
mounted on the rotation stage attached to the experiment table.
}
\end{figure}

\begin{figure}
\noindent\includegraphics[width=\textwidth]{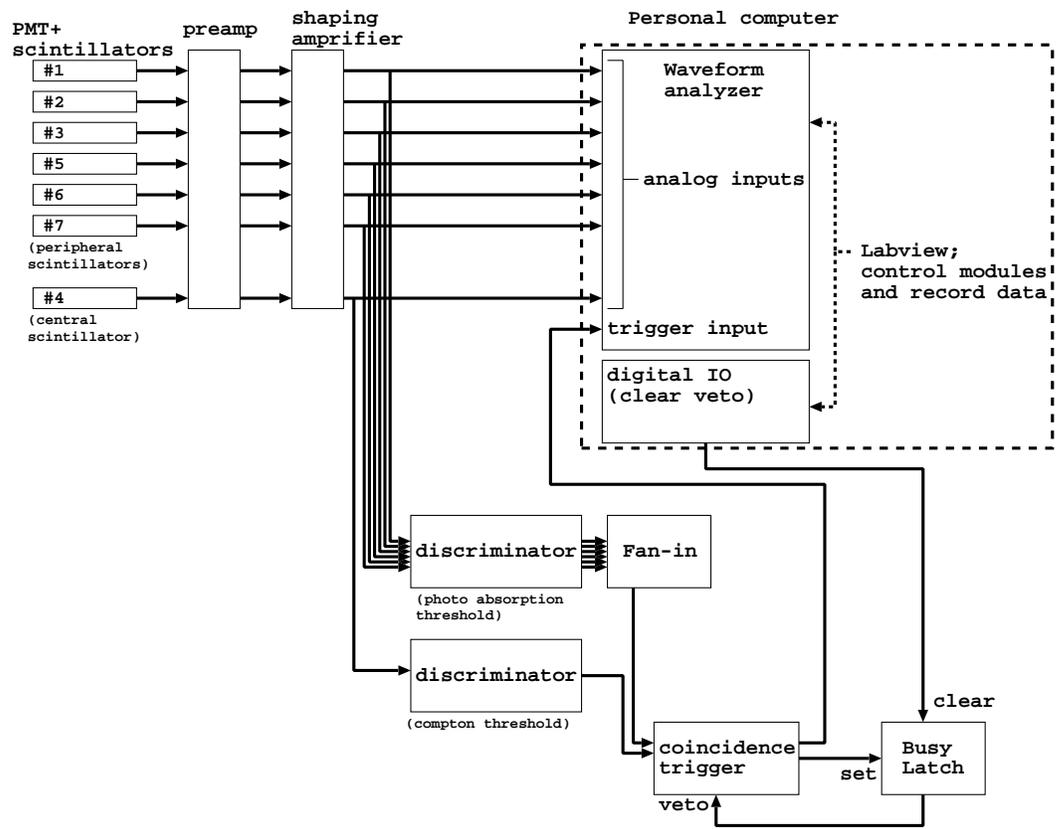}
\caption{
The data acquisition system of the experiment.
Outputs from shaping amplifiers are used for
trigger generation and spectroscopy;
the trigger was generated by a coincidence of hits
in the central scintillator and in any one of the peripheral
scintillators.
}
\end{figure}

\begin{figure}
\noindent\includegraphics[width=\textwidth]
{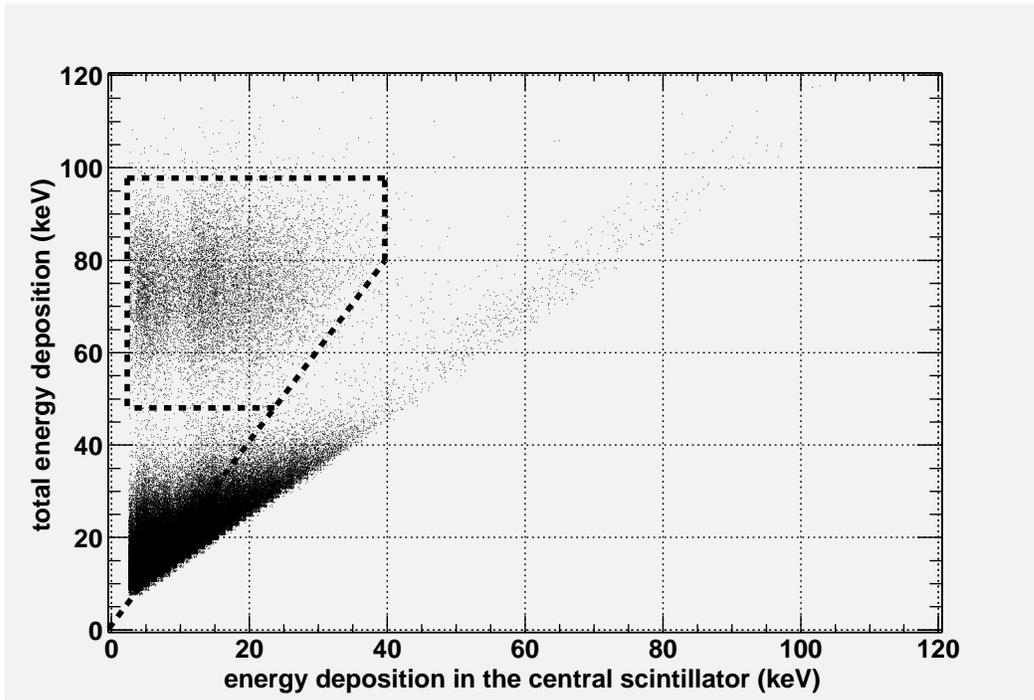}
\caption{
Relation of deposit energy in the central scintillator and 
total energy deposition for 73~keV run at 0-degree rotation.
Event selection criteria used in data analysis are also shown
by dotted lines. (see text)
}
\end{figure}

\begin{figure}
\noindent\includegraphics[width=\textwidth]{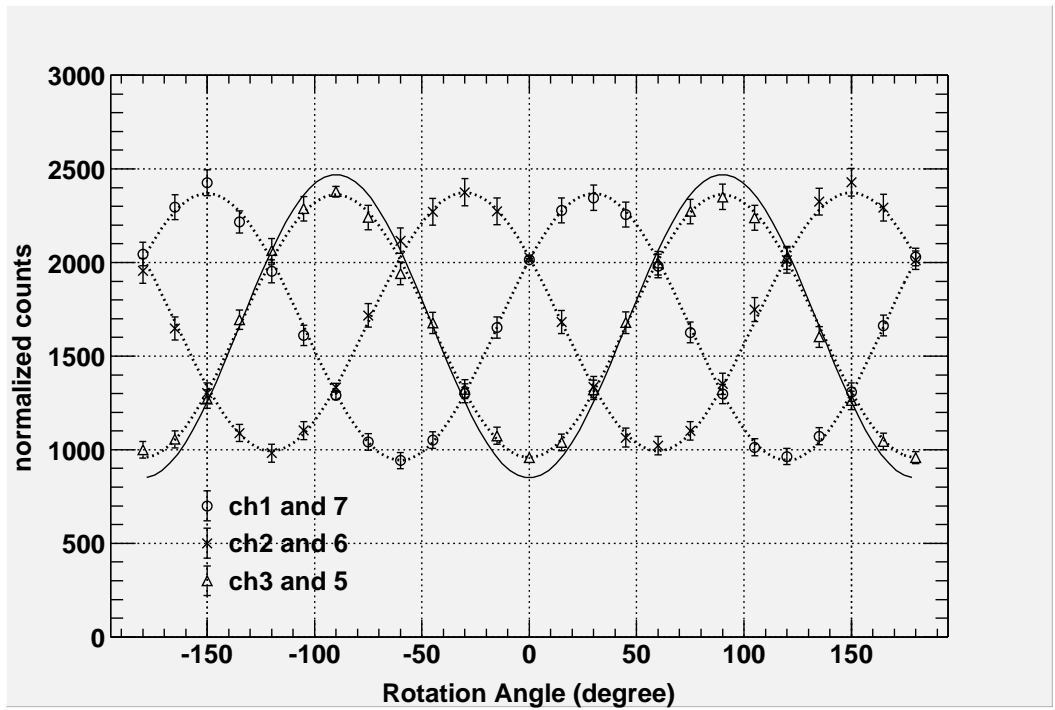}
\caption{
Normalized counts in peripheral scintillators as a function of
rotation angle for 73~keV run. Best fit models are
given as dotted lines.
For channels 3 and 5, the prediction by Geant4
(with modifications; see text) 
for 100\% linearly polarized beam
is shown by thin solid line.
}
\end{figure}

\clearpage

\appendix

\section{Geant4 Simulation of Polarized Photon Scattering}
\label{Geant4 Simulation of Polarized Photon Scattering}

Geant4 \cite{Geant4} is a toolkit for simulating the passage of
particles through matter. It has now become a
standard tool in a wide variety of fields,
e.g., high energy physics, medical science, astrophysics and 
space science.
It simulates a comprehensive range of physical processes
including electromagnetic, hadronic and optical interactions
and facilitates handling of complex geometries in the simulation.
It is, however, a relatively new product
(the first public release was in December 1998) and
needs to be validated by comparison with experimental data,
theoretical predictions and other simulation programs.

To reproduce our beam test data, we 
used Geant4 with the low energy extensions 
(G4LowEnergyPolarizedCompton class \cite{G4Pol}
and a G4LowEnergyRayleigh class) for photon scattering simulations.
We used version 5.1 and confirmed that
the classes had not been changed over versions
4.2--6.2 with regard to the polarized photon scattering.
Initial simulations gave a modulation factor of $\sim 0.37$,
implying an unphysical beam polarization of over 100\%.
We examined the Geant4 simulation program in detail
and found that even for the forward scattering case, 
in which energy transfer is negligible
and a scattered photon is expected to 
remain $\sim$ 100\% linearly polarized,
the polarization vector after the scattering changed
to some degree (Figure~A1).
We also noticed that the Rayleigh scattering 
extinguished the photon polarization vector,
and consequently gave an artificially small modulation.
We therefore had to modify the implementation of
polarized Compton/Rayleigh scattering in Geant4 as described below.

The notation is defined as in Figure~A2.
There, one completely linearly polarized photon is scattered 
by a free electron at point O.
The momentum vector of incident and scattered photon are
$\vec{k_{0}}$ and $\vec{k}$, respectively,
and the unit vector along the polarization vector 
before the scattering is $\vec{e_{0}}$.
$\theta$ and $\phi$ are the polar and azimuth angle of the scattering.
Here,
$\vec{k_{0}}$ and $\vec{e_{0}}$ are assumed to be along z-axis and
x-axis, respectively.
Then, the Klein-Nishina cross section of the Compton scattering is
given by
\begin{equation}
\frac{d\sigma}{d\Omega} = 
\frac{1}{2}{r_{0}}^{2} \frac{k^{2}}{{k_{0}}^2}
\left[
\frac{k}{k_{0}} + \frac{k_{0}}{k} -2 \sin^{2} \theta \cos^{2} \phi
\right]~~,
\end{equation}
and the Thomson scattering cross section is,
by substituting $k_{0}$ for $k$, given by
\begin{equation}
\frac{d\sigma}{d\Omega} = 
{r_{0}}^{2} 
\left(
1 - \sin^{2} \theta \cos^{2} \phi
\right)~~.
\end{equation}
We can see that photons are most likely scattered
at right angle to the direction of incident polarization vector
($\phi = 90^{\circ}$).
If we use the angle between 
the incident polarization vector ($\vec{e_{0}}$) and the
scattered polarization vector $\vec{e}$, the formula can 
also be expressed as
\begin{equation}
\frac{d\sigma}{d\Omega} = 
\frac{1}{4}{r_{0}}^{2} \frac{k^{2}}{{k_{0}}^2}
\left[
\frac{k}{k_{0}} + \frac{k_{0}}{k} -2+4 \cos^{4} \Theta
\right]~
\end{equation}
and
\begin{equation}
\frac{d\sigma}{d\Omega} = 
{r_{0}}^{2} \cos^{4} \Theta
\end{equation}
for the Compton and Thomson scattering,
respectively,
where $\Theta$ is the angle between two polarization 
vectors \cite{Heitler}.
The cross section per atom is obtained by taking into account the
incoherent scattering function (for the Compton scattering)
or the atomic form factor (for the Rayleigh scattering)
\cite{Hubbell1975,Hubbell1979}.

According to \cite{Heitler},
it is convenient to consider two directions for
$\vec{e}$:
one is in the same plane as $\vec{e_{0}}$ 
(denoted as $\vec{e_{\parallel}}$) and the other 
is perpendicular to it ($\vec{e_{\perp}}$).
Then, the differential cross section for these two directions is
\begin{equation}
\left( \frac{d\sigma}{d\Omega} \right)_{\parallel} = 
\frac{1}{4}{r_{0}}^{2} \frac{k^{2}}{{k_{0}}^2}
\left[
\frac{k}{k_{0}} + \frac{k_{0}}{k} -2
+4\left( 1-\sin^{2}\theta \cos^{2} \phi \right)
\right]~,
\end{equation}
and
\begin{equation}
\left( \frac{d\sigma}{d\Omega} \right)_{\perp} = 
\frac{1}{4}{r_{0}}^{2} \frac{k^{2}}{{k_{0}}^2}
\left[
\frac{k}{k_{0}} + \frac{k_{0}}{k} -2
\right]~~.
\end{equation}
The cross sections for the Thomson scattering is obtained by
substituting $k_{0}$ for $k$.
Then we obtain $\left( \frac{d\sigma}{d\Omega} \right)_{\perp}=0$ 
and can see that
the scattered photon is linearly polarized along $\vec{e_{\parallel}}$.
For the Compton scattering case, 
a scattered photon is  partially polarized along 
$\vec{e_{\parallel}}$ and the degree of polarization is calculated
from the maximum and minimum values of the cross section
($\left( \frac{d\sigma}{d\Omega} \right)_{\parallel}$ and
$\left( \frac{d\sigma}{d\Omega} \right)_{\perp}$, respectively) as
\cite{Rybicki}
\begin{equation}
P \equiv 
\frac{\left( \frac{d\sigma}{d\Omega} \right)_{\parallel} - \left( \frac{d\sigma}{d\Omega} \right)_{\perp}}
{\left( \frac{d\sigma}{d\Omega} \right)_{\parallel} + \left( \frac{d\sigma}{d\Omega} \right)_{\perp}}
= \frac{2\left( 1 - \sin^{2}\theta \cos^{2}\phi \right)}
{\frac{k}{k_{0}} + \frac{k_{0}}{k} -2 \sin^{2}\theta \cos^{2}\phi}~~.
\end{equation}
On the other hand, 
when the scattered photon is depolarized with a probability of $1-P$,
we sampled the direction of polarization vector at random in the plane
constructed by $\vec{e_{\parallel}}$ and $\vec{e_{\perp}}$.
Except for the way of sampling the polarization vector,
we used the original code for low-energy Compton/Rayleigh scattering.
The distribution of the polarization vector 
after Compton scattering in forward direction,
obtained by simulation with the new codes,
is also shown in Figure~A1.
We note that the way of sampling the polarization vector
is identical with that of EGS4 with polarized photon scattering
as described in \cite{Namito1993};
we have not implemented Doppler broadening 
yet \cite{Namito1994}.

We validated 
the Geant4 simulation with our modifications in detail
by comparing with the results of EGS4 simulation which included the effects
of Compton and Rayleigh scattering for polarized gamma-rays 
and the effect of Doppler broadening
\cite{Namito1993,Namito1994}.
The EGS4 program had been validated at the $\sim$ 10\% level
by comparing with polarization measurements \cite{Namito1993}.
In our validation test, we 
simulated a slab of 20~cm thick 
plastic scintillator and irradiated it with 5 million photons 
(100\% linearly polarized)
with a power-law energy spectrum
with an index of 2.1 to mimic the Crab Nebula spectrum
in 25--200~keV energy range \cite{Crab}.
Azimuthal angle asymmetry of the first and the second Compton scattering
are summarized in Figure~A3.
There, the distribution of Geant4 results with and without our modifications
are compared with EGS4 results.
Distributions of the first Compton scattering are almost identical
between EGS4 and the modified Geant4, giving a modulation factor of
$0.4941 \pm 0.0006$, whereas the original Geant4 shows a
smaller value of $0.4653 \pm 0.0006$ due to the (incorrect) 
Rayleigh scattering depolarization effect.
For the second Compton scattering,
EGS4 and the new Geant4 again give a consistent modulation factor
($0.3248 \pm 0.0008$).
On the other hand, the original Geant4 shows a much less
isotropic distribution
(modulation factor of only $0.1669 \pm 0.0008$)
due to the implementation problems already mentioned.

Finally, as an overall validation 
of polarized scattering processes,
we compared an expected modulation factor of the PoGO instrument
predicted by three simulation programs.
To do this 
we segmented hits in the plastic slab into
hexagonal elements of 2.68~cm thickness
(current design value of PoGO) and 
incorporated a typical PMT noise 
and the energy resolution of the scintillators.
The obtained modulation factor was 
$0.2184 \pm 0.0022$ and $0.2223 \pm 0.0023$ for EGS4 and 
modified Geant4, respectively, 
whereas that by original Geant4 was $0.1239 \pm 0.0023$
(Figure~A4).
We therefore conclude that the modified Geant4 and EGS4 
are consistent in 2--3\% level
when applied to simulating PoGO
with polarized Compton/Rayleigh scattering.
This study has already been reported to Geant4 team,
and new codes will be available in the public release
after the test is done.

\clearpage

\begin{figure}
\noindent\includegraphics[width=\textwidth]
{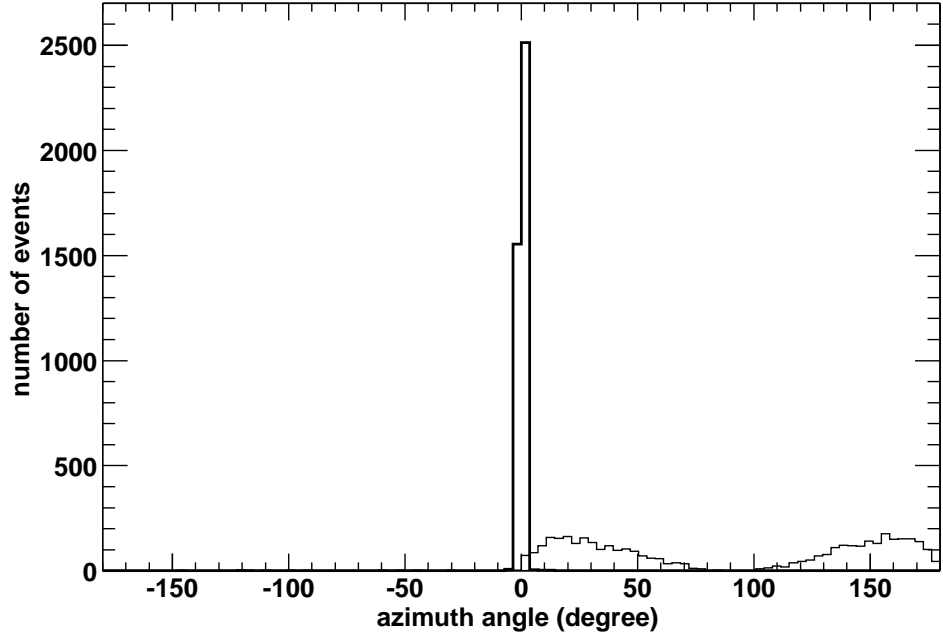}
\caption{
Azimuthal angle distribution of the polarization vector
of 100\% linearly polarized photons, 
after the first Compton scattering.
Beam energy is 100~keV and forward scattering events
($\cos \theta \ge 0.95$) are selected.
Distribution obtained by original Geant4 codes 
is given by thin solid line 
and that with our fixes by thick one.
The vector is parallel to the initial beam polarization vector
when the angle is $0^{\circ}$.
}
\end{figure}

\begin{figure}
\noindent\includegraphics[width=\textwidth]{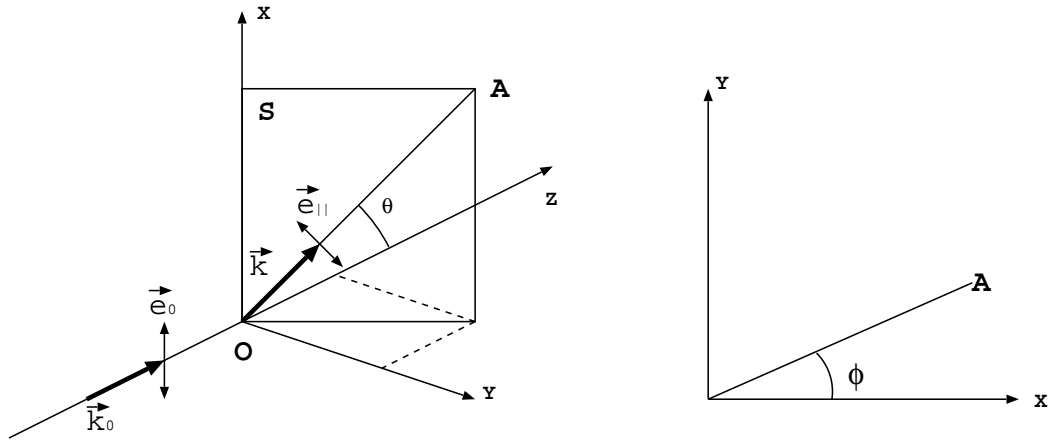}
\caption{
A photon scattering at point O.
The momentum vector ($\vec{k_{0}}$)
and the polarization vector ($\vec{e_{0}}$) of an incident photon
are along z- and x-axis, respectively.
$\theta$ and $\phi$ are the scattering polar and azimuth angle.
Plane S is constructed by $\vec{e_{0}}$ and $\vec{k}$,
the momentum vector after the scattering.
$\vec{e_{\parallel}}$ is in plane S.
}
\end{figure}

\begin{figure}
\noindent\includegraphics[width=\textwidth]{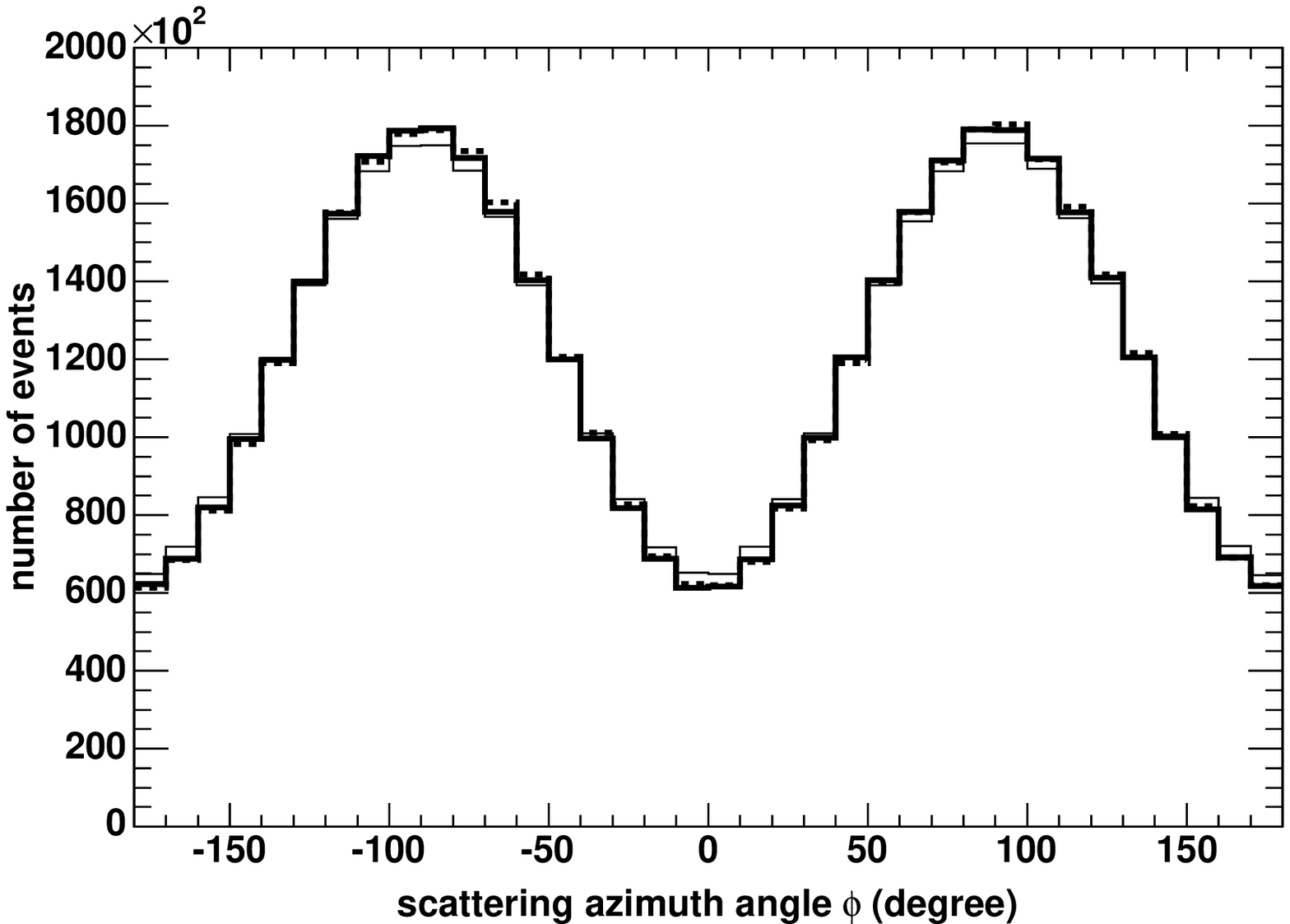}
\noindent\includegraphics[width=\textwidth]{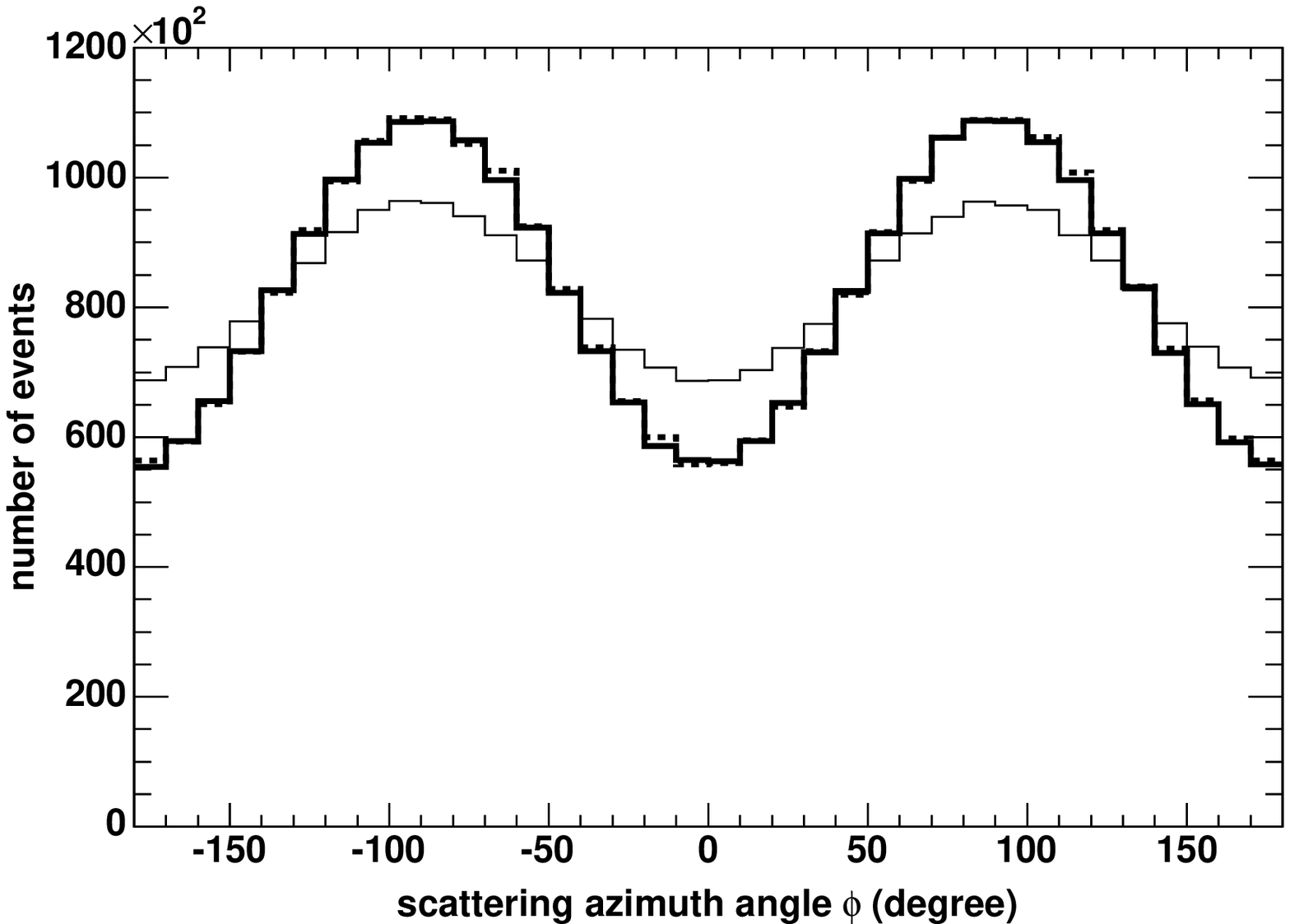}
\caption{
Azimuthal angle asymmetry of the first (top panel) 
and the second (bottom panel) Compton scattering for the 
fully polarized Crab spectrum in 25--200~keV,
where the angle is measured from the polarization vector
of the incident photon.
The distribution of modified Geant4, original Geant4 and EGS4 
(all with polarized Compton scattering process) are shown by 
thick solid line, thin solid one and dotted line, respectively.
For Geant4 with our fixes and EGS4 simulation,
polarized Rayleigh scattering is also implemented.
}
\end{figure}

\begin{figure}
\noindent\includegraphics[width=\textwidth]{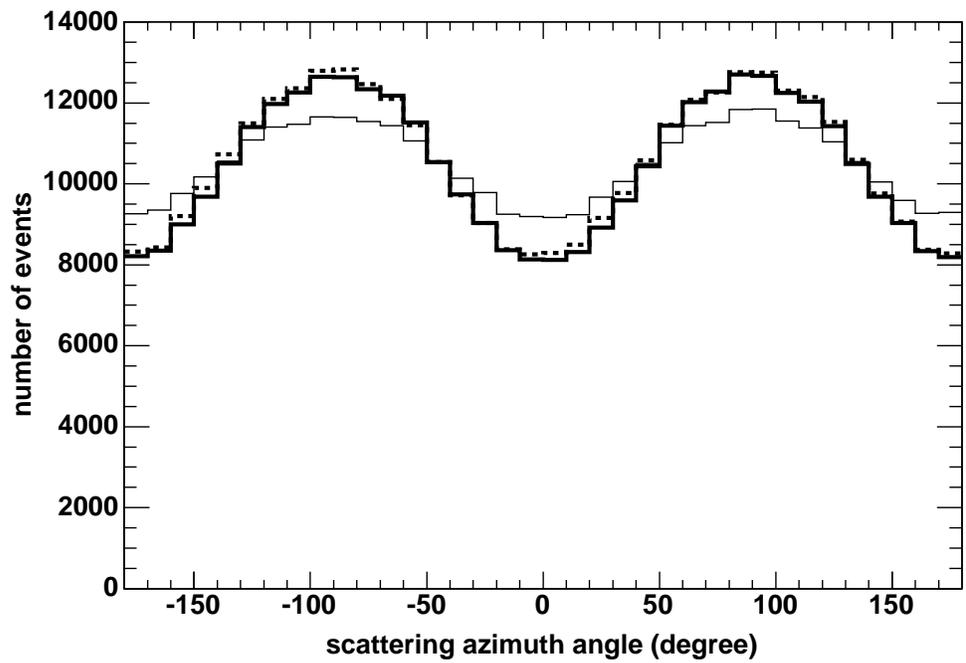}
\caption{
Predicted azimuthal angle distribution of the Crab Nebula spectrum
observed by the PoGO instrument.
Like for Figure~A3, the predictions by modified Geant4,
original Geant4 and EGS4 are shown by
thick solid line, thin solid one and dotted line,
respectively.
}
\end{figure}

\end{document}